\documentclass{article}
\usepackage{balance}
\usepackage{sectsty}
\usepackage{lastpage}
\usepackage{graphicx}
\usepackage{color}
\usepackage{epsfig}
\usepackage{amssymb}
\usepackage{caption}
\usepackage[cmex10]{amsmath}

\usepackage{array}
\begin{document}
%
\title{First-principles based simulations of electronic transmission in ReS$_{2}$/WSe$_{2}$ and ReS$_{2}$/MoSe$_{2}$ type-II vdW heterointerfaces}

\author{Dipankar~Saha\thanks{Department of Electrical Engineering, Indian Institute of Technology Bombay, Mumbai-400076, India.} \thanks{Department of Electronics and Telecommunication Engineering, Indian Institute of Engineering Science and Technology Shibpur, Howrah-711103, India.} \thanks{email: dipsah\textunderscore etc@yahoo.co.in} and~Saurabh~Lodha \footnotemark[1] 
}
\date{}
\maketitle

\begin{abstract}
Electronic transmission in monolayer ReS$_{2}$ and ReS$_{2}$ based van der Waals (vdW) heterointerfaces are studied here. Since ReS$_{2}$/WSe$_{2}$ and ReS$_{2}$/MoSe$_{2}$ type-II vdW heterostructures are suitable for near infrared (NIR)/short-wave infrared (SWIR) photodetection, the role of interlayer coupling at the heterointerfaces is examined in this work. Besides, a detailed theoretical study is presented employing density functional theory (DFT) and nonequilibrium Green’s function (NEGF) combination to analyse the transmission spectra of the two-port devices with ReS$_{2}$/WSe$_{2}$ and ReS$_{2}$/MoSe$_{2}$ channels and compare the near-equilibrium conductance values. 
\\
Single layer distorted 1T ReS$_{2}$ exhibits formation of parallel chains of \textquoteleft Re\textquoteright-\textquoteleft Re\textquoteright$\,$bonds, leading to in-plane anisotropy. Owing to this structural anisotropy, the charge carrier transport is very much orientation dependent in ReS$_{2}$. Therefore, this work is further extended to investigate the role of clusterized \textquoteleft Re\textquoteright$\,$ atoms in electronic transmission.
\end{abstract}


%

\section{Introduction}

In recent years significant progress in micro and nano photonics has been observed owing to the optoelectronic devices based on atomically thin two-dimensional (2D) layered materials and their vertical stacks \cite{Koppens,TimothyAnnuRev,ITRS,AndrasKisPhotodetector, ThorstenDeilmann,KartikeyThakar,SayantanGhosh,AKGeimGrigorieva,PhilipKim,MarcoM.Furchi,PulickelAjayan,ThomasHeine, ChenhaoJin,Akinwande}. Such 2D materials can be metallic (graphene) or, semiconducting (e.g., black phosphorus, transition metal dichalcogenides (TMDs), etc.) or, a combination of both \cite{AvourisGraphene,QGuoBlackPhosphorus,Matheiu Massicotte,AlamriWS2Graphene,ChulhoPark_ReS2WSe2Nanoscale}. The vdW heterointerfaces formed with vertical stacking of 2D materials may exhibit compelling new properties which are significantly different from those of the participant materials \cite{AKGeimGrigorieva,DSahaACSanm,PranjalKumarGogoi}. Strong interlayer coupling and fast charge transfer across vdW interface are the key features which largely determine performance efficiency of the heterostructures \cite{AKGeimGrigorieva,PhilipKim,DSahaACSanm,AbinNanoLett,DinhHoaLuong}. In oder to design ultrafast NIR and SWIR photodetectors, a theoretical study exploring various possible combinations of group-6 and group-7 monolayer TMDs is presented in \cite{DSahaACSanm}. Among different combinations of type-II vdW heterostructures, it has been found that both ReS$_{2}$/WSe$_{2}$ and ReS$_{2}$/MoSe$_{2}$ emerge as suitable candidates with efficient generation, separation, and collection of charge carriers \cite{DSahaACSanm}. Howbeit, the effects of interlayer coupling on the electronic structures of ReS$_{2}$/WSe$_{2}$ and ReS$_{2}$/MoSe$_{2}$ were not captured in \cite{DSahaACSanm}. Moreover, to quantify the near-equilibrium conductance values of those heterointerfaces, a detailed analysis of electronic transmission through the ReS$_{2}$/WSe$_{2}$ and the ReS$_{2}$/MoSe$_{2}$ needs to be conducted.
\\
Thus, in this work, considering band dispersions and electron difference density (EDD) calculations, first we try to investigate the role of interlayer coupling at the vdW heterojunctions. Next, utilizing NEGF along with DFT, we have shown the electronic transmission of two-port devices with the ReS$_{2}$/WSe$_{2}$ and the ReS$_{2}$/MoSe$_{2}$ channels and compared their conductance values at near-equilibrium.  
\\

Besides, we have also investigated the anisotropic carrier transmission of the group-7 constituent material, that is distorted 1T ReS$_{2}$. But for the group-6 TMDs (e.g., MoS$_{2}$, WS$_{2}$, MoSe$_{2}$, and WSe$_{2}$), because of the symmetric honeycomb structure of semiconducting 2H phase, such in-plane anisotropy is generally not expected \cite{DSahaACSanm,DipankarPCCP}. Furthermore, it is worth mentioning that orientation dependent electro-thermal transport in 2D layered materials can be useful for the purpose of designing low power, ultrathin next generation electronic devices \cite{DipankarPCCP,YCLin,DanielAChenet,DavidGCahill, NGoyalACSami,MohammadRahman}. Among various 2D materials, in-plane anisotropy owing to structural transformation is seen in the 1T$^{\prime}$ phase of the MoS$_{2}$ \cite{DipankarPCCP,ManishChhowallaNchem}. Strong anisotropic conductance, due to clusterization of \textquoteleft Mo\textquoteright$\,$ atoms along the transport direction, is reported in \cite{DipankarPCCP}. A similar trend can be seen in distorted 1T ReS$_{2}$, where the formation of parallel chains of \textquoteleft Re\textquoteright-\textquoteleft Re\textquoteright$\,$bonds can lead to orientation dependent anisotropic transport \cite{YCLin}. As reported in \cite{YCLin}, distorted 1T ReS$_{2}$ exhibits direction dependent I-V as well as transfer characteristics. Such experimental observation motivated us to explore the electronic properties of single layer distorted 1T ReS$_{2}$ and investigate the role of clusterized \textquoteleft Re\textquoteright$\,$ atoms in carrier transmission.

\section{Methodology}
In order to coduct first-principles based DFT calculations, the software package \textquotedblleft QuantumATK\textquotedblright$\,$ was used \cite{ATK201806,newATKRefIOP}. For all the unit cells and the supercells, computation of electronic structures and geometry optimizations were performed using the generalized gradient approximation (GGA) as exchange correlation along with the Perdew-Burke-Ernzerhof (PBE) functional \cite{PBE}. Moreover, LCAO (linear combination of atomic orbitals) based numerical basis sets were utilized in this study to obtain results at the cost of reasonable computational load \cite{DipankarPCCP,ATK201806}. To attain good accuracy of quantum transport and electronic structure calculations, the OPENMX (Open source package for Material eXplorer) code was used as the norm-conserving pseudopotentials \cite{OMX_1,OMX_2}. The basis sets for Mo, S, W, Se, and Re were taken as \textquotedblleft s3p2d1\textquotedblright, \textquotedblleft s2p2d1\textquotedblright, \textquotedblleft s3p2d1\textquotedblright, \textquotedblleft s2p2d1\textquotedblright, and \textquotedblleft s3p2d1\textquotedblright$\,$ respectively. Besides, the density mesh cut-off value was set to 200 Hartree and the k-points in Monkhorst-Pack grid were set to 9$\times$9$\times$1 (X-Y-Z) for the unit cells and 9$\times$9$\times$3 for the heterostructures. Sufficient vacuum was incorporated along the direction normal to the in-plane for the purpose of avoiding spurious interaction between periodic images. Furthermore, to include the effects of vdW interactions among different monolayers, the Grimme's dispersion correction (DFTD2) was employed  \cite{Grimme}. 
\\
Considering the two-port devices, transmission spectra along the channels were computed utilizing NEGF along with DFT. The k-points in Monkhorst-Pack grid were adopted as 1$\times$9$\times$150 for the device calculations. Apart from that, in order to solve the Poisson's equation, Dirichlet boundary condition in the transport direction (Z direction) and periodic boundary conditions in the other two directions (X–Y directions) were assigned. In the framework of NEGF, taking into account the broadening matrices $\Gamma_{l,r}(E)=i[\Sigma_{l,r}-\Sigma_{l,r}^{\dagger}]$ for the left and the right electrodes (set by computing the self energy matrices $\Sigma_{l}$ and $\Sigma_{r}$) \cite{Markussen,D_Saha_JAP}, the electronic transmission is calculated as,        
\begin{equation} 
T_{e}(E)=\textnormal{Tr}[\Gamma_{l}(E)\: G(E)\:\Gamma_{r}(E)\: G^{\dagger}(E)]\;, 
\end{equation} where $G(E)$ and $G^{\dagger}(E)$ denote retarded Green's function and advanced Green's function respectively \cite{DipankarPCCP,Markussen,Brandbyge}. 

\section{Results and Discussion}

Figure 1 illustrates a typical type-II vdW heterostructure formed with vertical stacking of group-6 and group-7 monolayer TMDs, as well as highlights its salient features \cite{DSahaACSanm}. Considering NIR/SWIR photodetection, near-direct bandgap, larger band offset values, and feasibility of integrating with flexible substrates make monolayer ReS$_{2}$ (group-7 TMD)/ monolayer WSe$_{2}$ (group-6 TMD) heterointerface a potentially promising candidate for the next generation ultrathin optoelectronic device \cite{TimothyAnnuRev,PhilipKim,Matheiu Massicotte,DSahaACSanm}. Even with the increase in number of layers of ReS$_{2}$ and WSe$_{2}$, we find that the near-direct type-II band alignment remains unaltered, hence ensuring efficient optical generation across the vdW interface \cite{AbinNanoLett}. Apart from that, monolayer ReS$_{2}$ (group-7 TMD)/ monolayer MoSe$_{2}$ (group-6 TMD) heterostructure can also be suitable for NIR/SWIR photodetection owing to its near-direct bandgap and superior optical absorption of the constituent MoSe$_{2}$ layer \cite{DSahaACSanm}. 
\\

\begin{figure}[!htbp]
\begin{center}
\includegraphics [height=1.843in,width=3.2in]{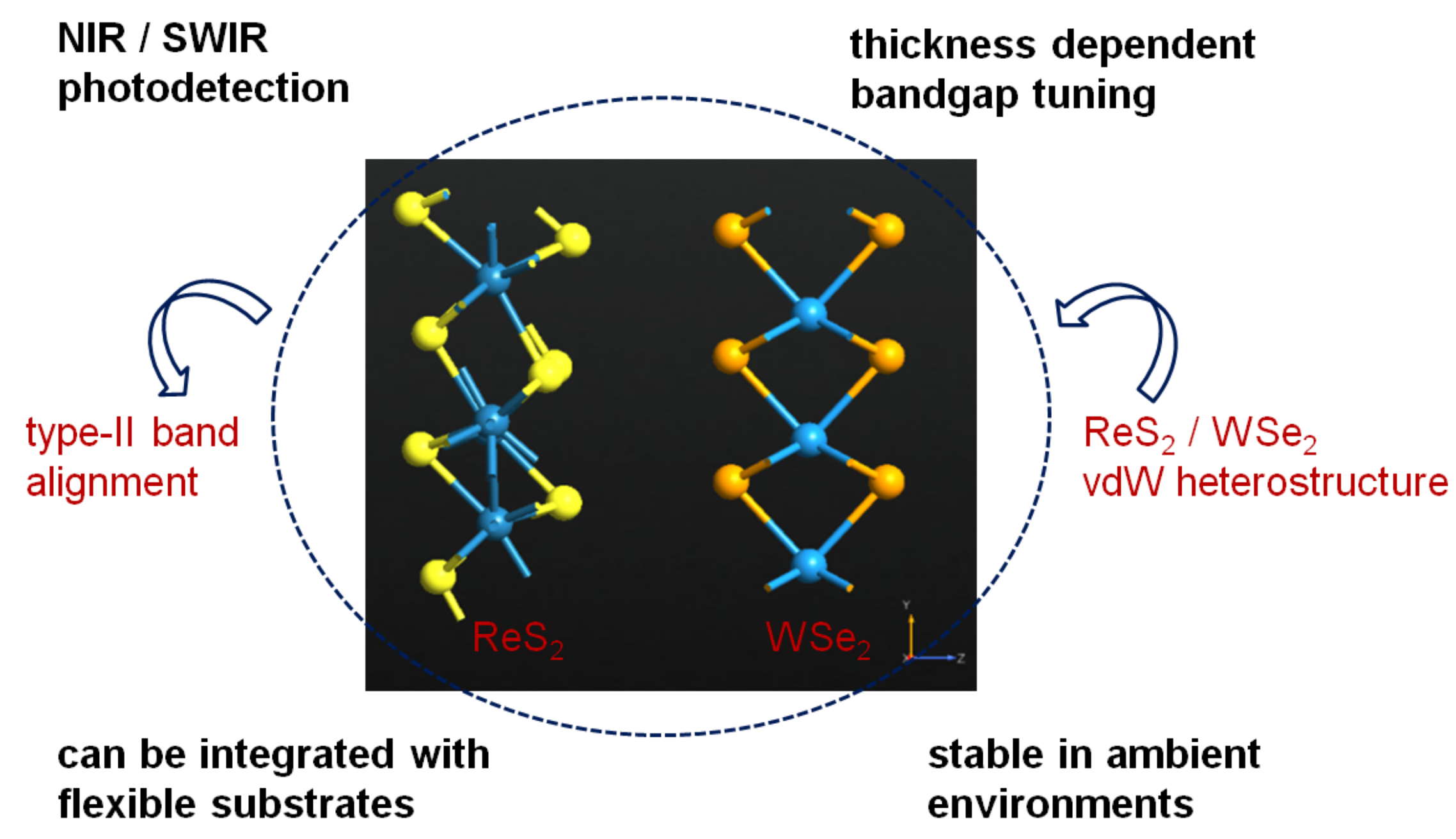}
\caption{Vertical stacking of group-6 and group-7 monolayer TMDs to form a type-II vdW heterointerface (viz. ReS$_{2}$/WSe$_{2}$).} 
\end{center}
\label{Figure1}
\end{figure}
Figure 2 and 3 depict the electronic structures of ReS$_{2}$/WSe$_{2}$ and ReS$_{2}$/MoSe$_{2}$ considering element-wise contributions \cite{Wyckoff,W_J_Schutte,KristinPersson}. Both the projected density of states (DOS) plots exhibit type-II band alignments with energy gap values of 0.684 eV and 0.842 eV for the ReS$_{2}$/WSe$_{2}$ and the ReS$_{2}$/MoSe$_{2}$ heterostructures. Besides, the in-plane lattice constants and the equilibrium interlayer distance for the hexagonal vdW heterointerface of ReS$_{2}$/WSe$_{2}$ (Fig. 2 inset) are a=b= 6.600 \AA$\;$ and 3.49 \AA$\;$ respectively \cite{DSahaACSanm}. Those values for ReS$_{2}$/MoSe$_{2}$ interface (Fig. 3 inset) are a=b= 6.597 \AA$\;$ and 3.41 \AA. \cite{DSahaACSanm}.    
\begin{figure}[!htbp]
\begin{center}
\includegraphics [height=2.862in,width=2.8in]{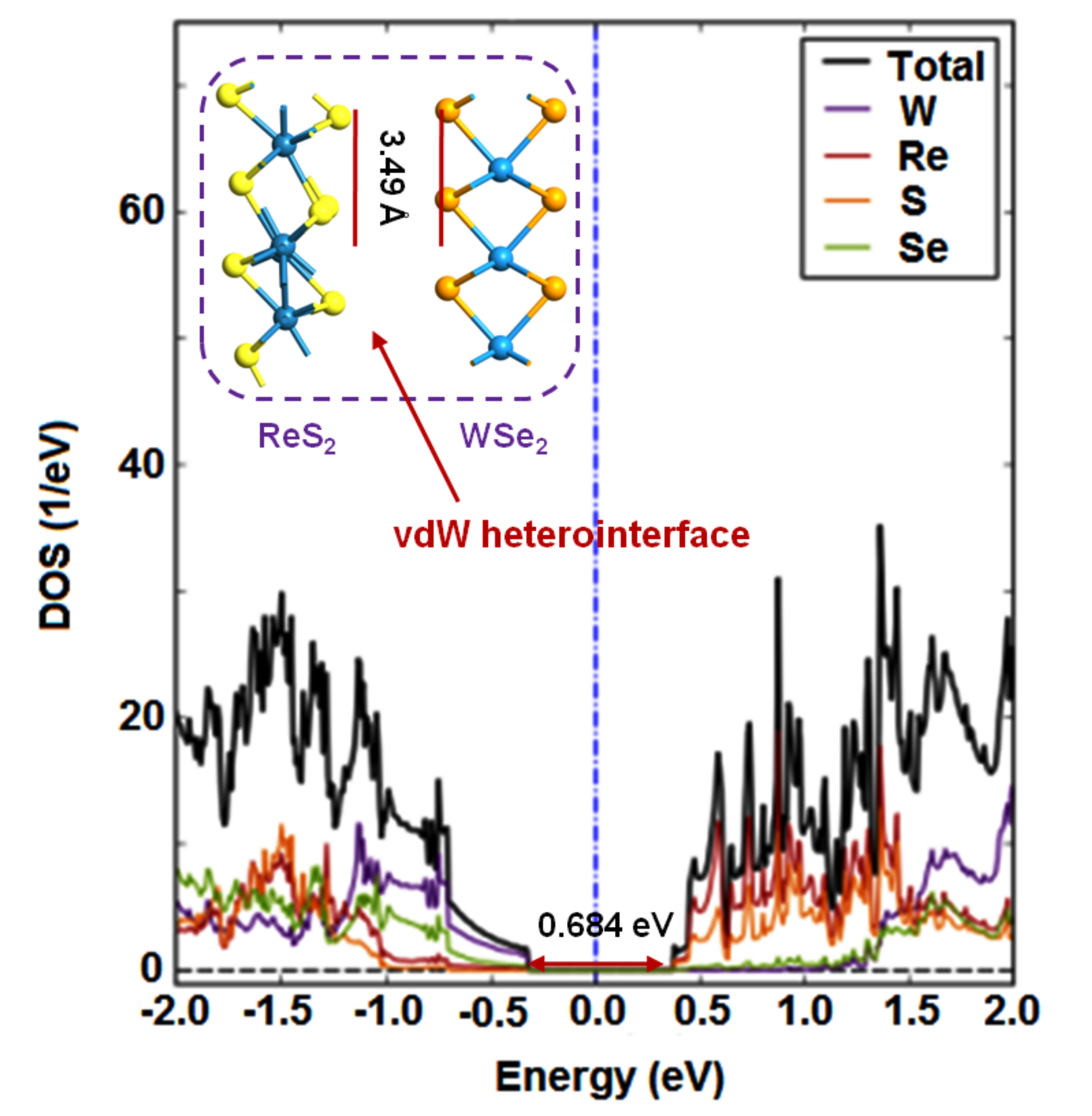}
\caption{Projected DOS plots for ReS$_{2}$/WSe$_{2}$ vdW heterointerface.} 
\end{center}
\label{Figure2}
\end{figure}

\begin{figure}[!htbp]
\begin{center}
\includegraphics [height=3.04in,width=2.8in]{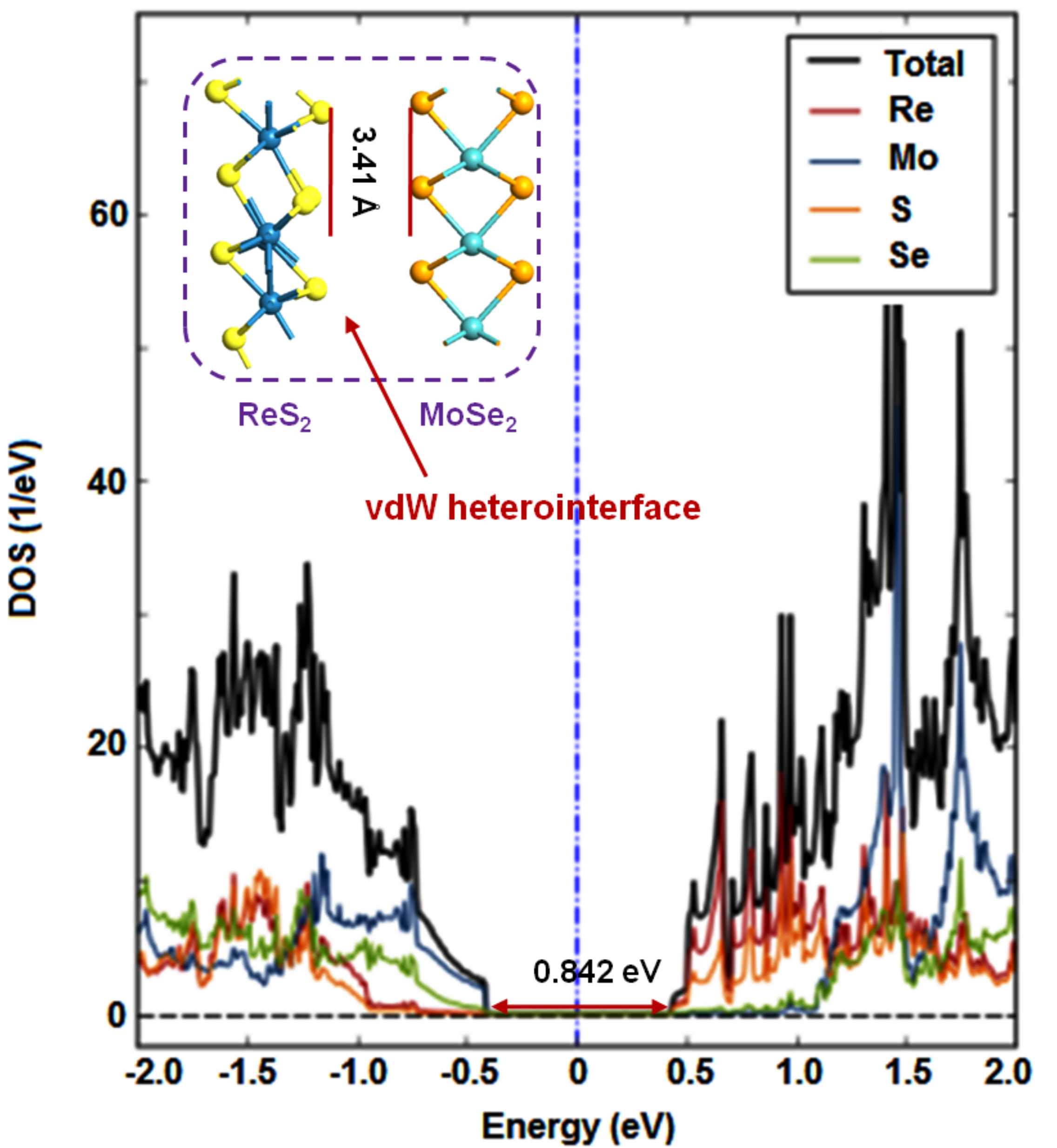}
\caption{Projected DOS plots for ReS$_{2}$/MoSe$_{2}$ vdW heterointerface.} 
\end{center}
\label{Figure3}
\end{figure}

In order to determine the effect of interlayer coupling, next we carry out a detailed analysis which emphasizes mainly on the band dispersions and the electron difference density (EDD) calculations of the heterointerfaces \cite{ShiSSTRev}. We consider the equilibrium interlayer distances of ReS$_{2}$/WSe$_{2}$ and ReS$_{2}$/MoSe$_{2}$ as the references (which are computed while optimizing the geometries utilizing LBFGS (Limited-memory Broyden Fletcher Goldfarb Shanno) algorithm). We also maintain the stacking patterns of the fully relaxed heterostructures same as those obtained form geometry optimization calculations. We then vary the interlayer distances (d\textsubscript{int}), and compute band dispersions of the vdW heterostructures (Fig. S1(a) and S1(b)). The changes in indirect and direct bandgap (I\textsubscript{gap} and D\textsubscript{gap}) values, as we deviate from the equilibrium interlayer distances, are listed in Table 1. 

I\textsubscript{gap} and D\textsubscript{gap} for ReS$_{2}$/WSe$_{2}$ are 0.68 eV and 0.83 eV, when the d\textsubscript{int} is 3.49 \AA$\;$ (equilibrium interlayer distance). On the other hand, I\textsubscript{gap} and D\textsubscript{gap} for ReS$_{2}$/MoSe$_{2}$ are 0.84 eV and 0.98 eV, when the d\textsubscript{int} is 3.41 \AA$\;$ (equilibrium interlayer distance). Now, as we shift the d\textsubscript{int} around the equilibrium  distances, we find that there are no noteworthy changes in band dispersions of ReS$_{2}$/WSe$_{2}$ and ReS$_{2}$/MoSe$_{2}$ (Fig. S1(a) and S1(b)), though the I\textsubscript{gap} and the D\textsubscript{gap} values do vary. We denote these changes as $|$$\Delta$E\textsubscript{I}$|$ and $|$$\Delta$E\textsubscript{D}$|$ (Table 1). Moreover, it is important to note that the variations in $|$$\Delta$E\textsubscript{I}$|$ and $|$$\Delta$E\textsubscript{D}$|$ are similar for the aforementioned type-II heterostructures. 
\\

Hence, we look into another important aspect, that is the EDD or the charge re-distribution between the constituent layers. As depicted in the cut plane diagrams of Fig. 4, negative and positive values of EDD indicate charge depletion and charge accumulation respectively. The EDD plot for ReS$_{2}$/WSe$_{2}$ (Fig. 4(a)) delineates minimum and maximum values of difference density as -0.29 and 0.2 \AA$^{-3}$. Thus the corresponding average value is $\sim$ -0.04 \AA$^{-3}$. On the other hand, EDD plot for ReS$_{2}$/MoSe$_{2}$ (Fig. 4(b)) shows minimum and maximum values of difference density as -0.29 and 0.79 \AA$^{-3}$, with the average of $\sim$ 0.25 \AA$^{-3}$. The larger average value of EDD for the ReS$_{2}$/MoSe$_{2}$ interface is essentially interpreting more charge distribution. Therefore, we can expect much stronger effect of interlayer coupling at the heterointerface of ReS$_{2}$/MoSe$_{2}$. In the following, we have further compared the electronic transmission in both the vdW heterointerfaces and observed a similar trend.
\\

\begin{table}[!t]
\renewcommand{\arraystretch}{1.5}
\caption{Change in bandgap with the varying interlayer distances}
\label{coupling}
\centering
\begin{tabular} {c | c  c  c  c  c }
\hline
\bfseries interface &  \bfseries d\textsubscript{int} (\AA) & \bfseries I\textsubscript{gap} (eV) & \bfseries D\textsubscript{gap} (eV) & \bfseries $|$$\Delta$E\textsubscript{I}$|$ (eV) & \bfseries $|$$\Delta$E\textsubscript{D}$|$ (eV) \\
\hline
 & 2.69 & 0.80 & 0.89 & 0.12 & 0.06 \\  
ReS$_{2}$/WSe$_{2}$ & 3.49 & 0.68 & 0.83 & 0.00 & 0.00 \\ 
 & 4.30 & 0.66 & 0.83 & 0.02 & 0.00  \\ 
 & 5.10 & 0.66 & 0.82 & 0.02 & 0.01 \\ 
\hline
 & 2.71 & 0.92 & 0.96 & 0.08 & 0.02 \\
ReS$_{2}$/MoSe$_{2}$ & 3.41 & 0.84 & 0.98 & 0.00 & 0.00 \\
 & 4.48 & 0.82 & 0.98 & 0.02 & 0.00 \\
 & 5.18 & 0.81 & 0.97 & 0.03 & 0.01 \\
\hline
\end{tabular}
\end{table}

\begin{figure}[!htbp]
\begin{center}
\includegraphics [height=4.302in,width=3in]{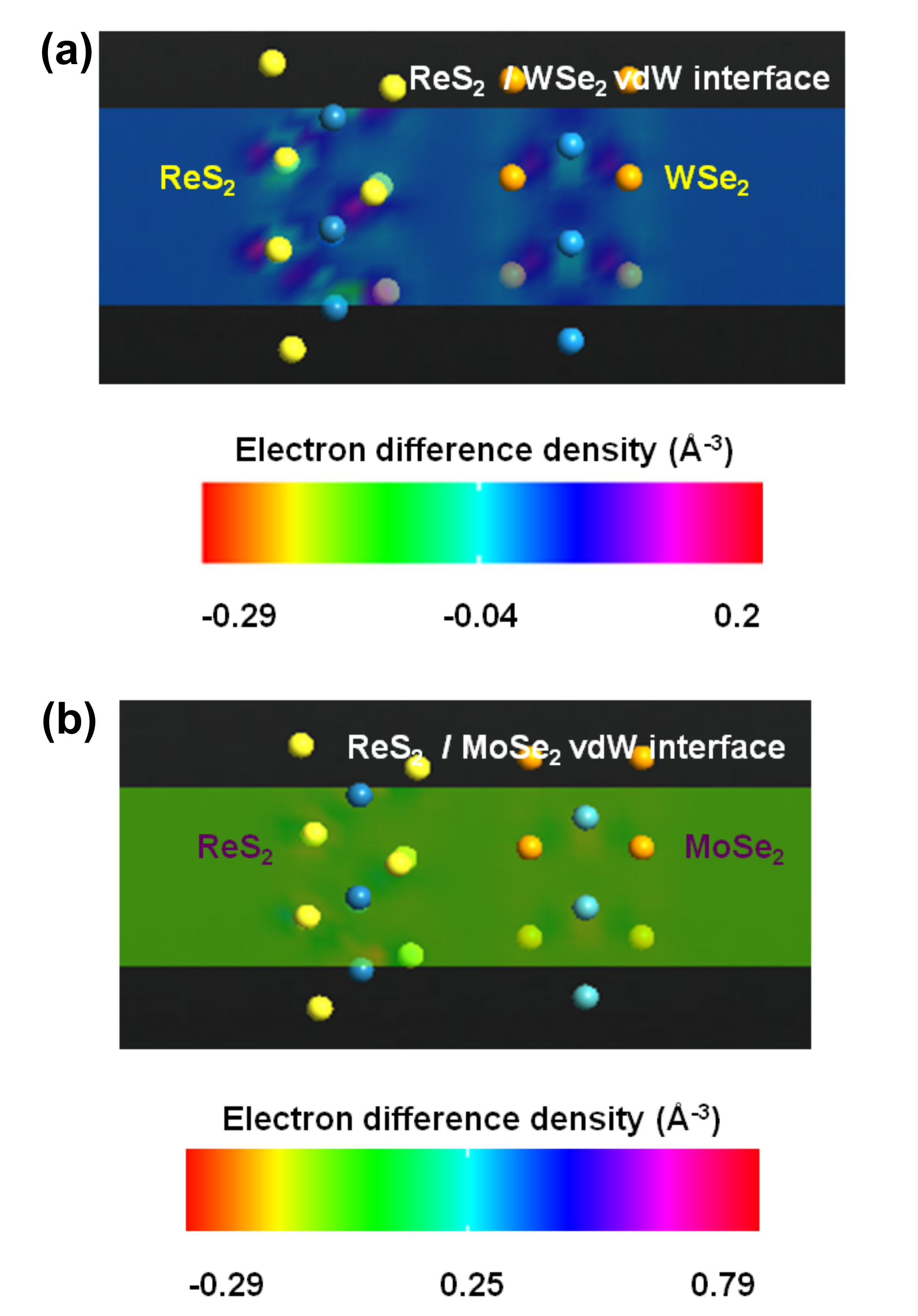}
\caption{EDD plots showing charge re-distribution between the constituent
layers of (a) ReS$_{2}$/WSe$_{2}$ and (b) ReS$_{2}$/MoSe$_{2}$.} 
\end{center}
\label{Figure4}
\end{figure}

Next, to model the two-port devices, we take into account the geometry optimized vertical stacks of ReS$_{2}$/WSe$_{2}$ and ReS$_{2}$/MoSe$_{2}$ as illustrated in Fig. 2 and Fig. 3 (insets). As shown in Fig. 5, the group-7 and the group-6 TMD layers are extended to form the left and the right electrodes respectively. For the ReS$_{2}$/WSe$_{2}$ device, t\textsubscript{left} (length of the left electrode) = t\textsubscript{right} (length of the right electrode) is $\sim$ 6.60 \AA. For the ReS$_{2}$/MoSe$_{2}$ device, that value is $\sim$ 6.59 \AA. The length and the width of the channel region of ReS$_{2}$/WSe$_{2}$ two-port device are 6.658 nm and 1.143 nm (Fig. 5(a)). Considering the channel region of ReS$_{2}$/MoSe$_{2}$ two-port device, those values are 6.654 nm and 1.142 nm respectively (Fig. 5(b)). Besides, for those composite two-port device structures, the overlapping regions are maintained as $\sim$ 2.423 nm. It is worth mentioning that the mean absolute strain values on both the monolayer surfaces are 0.61\% and 0.59\% for ReS$_{2}$/WSe$_{2}$ and ReS$_{2}$/MoSe$_{2}$ channels \cite{DSahaACSanm}. 
\\

\begin{figure}[!htbp]
\begin{center}
\includegraphics [height=1.883in,width=3.2in]{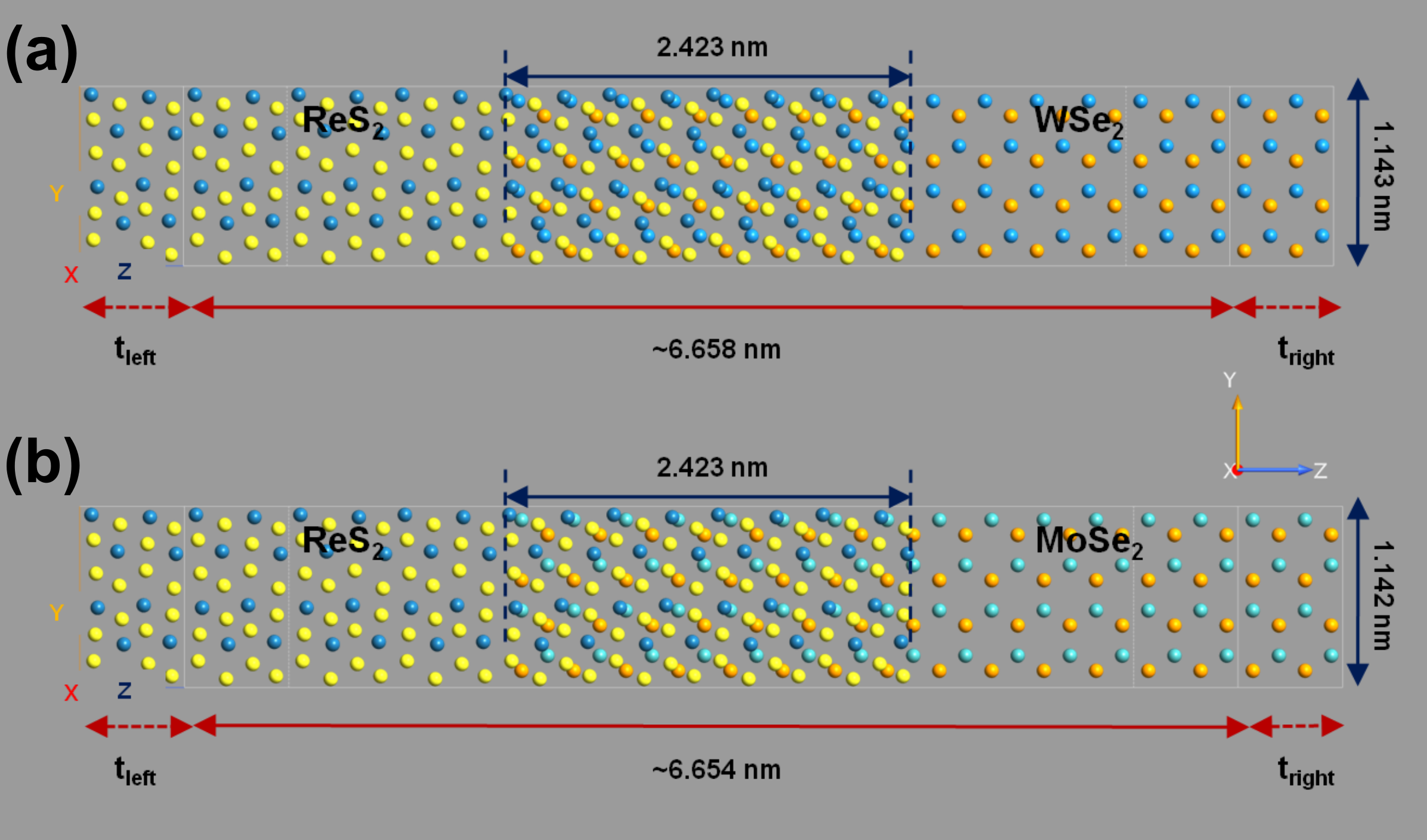}
\caption{Atomistic models of the two-port devices with (a) ReS$_{2}$/WSe$_{2}$ and (b) ReS$_{2}$/MoSe$_{2}$ channels.} 
\end{center}
\label{Figure5}
\end{figure}

\begin{figure}[!htbp]
\begin{center}
\includegraphics [height=1.717in,width=3.4in]{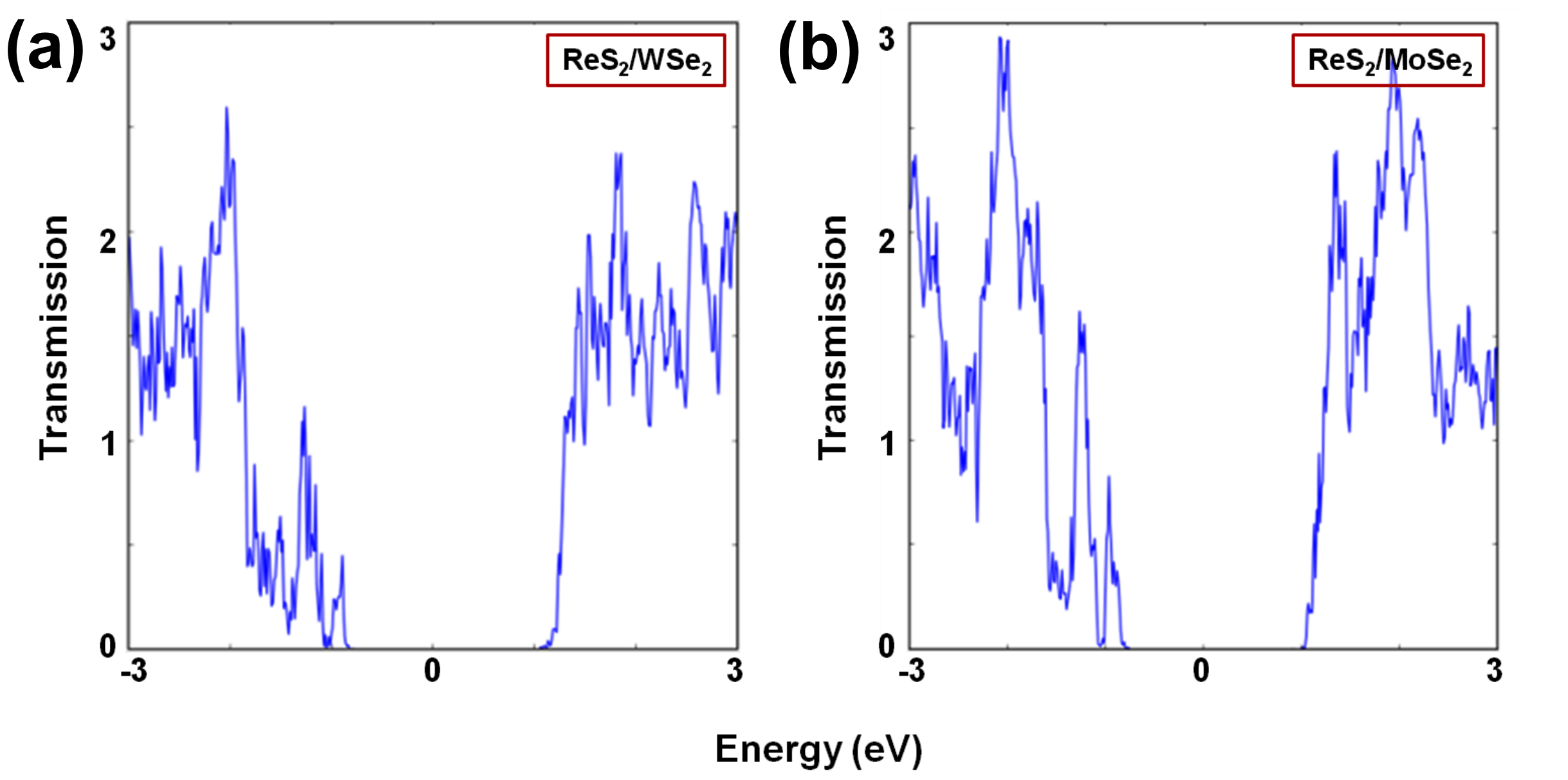}
\caption{Transmission spectra of the two-port devices with (a) ReS$_{2}$/WSe$_{2}$ and (b) ReS$_{2}$/MoSe$_{2}$ channels.} 
\end{center}
\label{Figure6}
\end{figure}
Figure 6 shows the transmission spectra (zero bias) of the two-port devices, where we can observe that the transmission through ReS$_{2}$/MoSe$_{2}$ channel is slightly better than that in ReS$_{2}$/WSe$_{2}$ (cosidering a small energy range near the energy zero level). In order to quantify this we further calculate the near-equilibrium conductance and Seebeck coefficient values, as listed in Table 2. S and G\textsubscript{e} denote the Seebeck coefficient and the electrical conductance, computed utilizing the linear response approximation ($\sim$ 300 K) \cite{DipankarPCCP,D_Saha_JAP}. $\Delta$E\textsubscript{F} represents the shift in energy level from the energy zero. Moreover, the plots of Seebeck coefficients for the energy range of -3 eV to 3 eV are shown in Fig. S2. Similar to the trends of transmission spectra (Fig. 6), the G\textsubscript{e} values listed Table 2 reinforce that the ReS$_{2}$/MoSe$_{2}$ channel is more conducive to the charge carrier transport. 
  
\begin{table}[!t]
\renewcommand{\arraystretch}{1.5}
\caption{Details of near-equilibrium conductance values}
\label{conductance}
\centering
\begin{tabular} {c | c  c  c }
\hline
\bfseries two-port devices & \bfseries $\Delta$E\textsubscript{F} (eV) & \bfseries G\textsubscript{e} (S) & \bfseries S (V/K) \\
\hline
 & 0.9 & 2.906$\times$10\textsuperscript{-9} & -0.0004686 \\
 & 1.0 & 9.022$\times$10\textsuperscript{-8} & -0.0002988 \\
ReS$_{2}$/WSe$_{2}$ & 1.1 & 1.023$\times$10\textsuperscript{-6} & -0.0001835 \\
(length = 6.658 nm & 1.2 & 8.563$\times$10\textsuperscript{-6} & -0.0001511 \\
width = 1.143 nm) & 1.3 & 3.540$\times$10\textsuperscript{-5} & -4.829$\times$10\textsuperscript{-5} \\
 & 1.4 & 5.180$\times$10\textsuperscript{-5} & -1.856$\times$10\textsuperscript{-5} \\
 & 1.8 & 7.897$\times$10\textsuperscript{-5} & -9.423$\times$10\textsuperscript{-6} \\
\hline
 & 0.9 & 4.925$\times$10\textsuperscript{-8} & -0.0004324 \\
 & 1.0 & 1.396$\times$10\textsuperscript{-6} & -0.0002594 \\
ReS$_{2}$/MoSe$_{2}$ & 1.1 & 1.252$\times$10\textsuperscript{-5} & -0.000103 \\
(length = 6.654 nm & 1.2 & 3.497$\times$10\textsuperscript{-5} & -5.931$\times$10\textsuperscript{-5} \\
width = 1.142 nm) & 1.3 & 6.848$\times$10\textsuperscript{-5} & -2.906$\times$10\textsuperscript{-5} \\
 & 1.4 & 7.434$\times$10\textsuperscript{-5} & -1.313$\times$10\textsuperscript{-5} \\
 & 1.8 & 8.396$\times$10\textsuperscript{-5} & -1.164$\times$10\textsuperscript{-5} \\
\hline
\end{tabular}
\end{table}

Next, we have extended this theoretical study to demonstrate in-plane anisotropy and orientation dependant carrier transmission in single layer distorted 1T ReS$_{2}$. For the purpose of modeling the two-port devices with distorted 1T ReS$_{2}$ channels, we consider three different supercells viz. \textquoteleft ReS$_{2}$ supercell1\textquoteright$\,$(where the diamond-shaped (DS) \textquoteleft Re\textquoteright$\,$chains are perpendicular to the transport direction), \textquoteleft ReS$_{2}$ supercell2\textquoteright$\,$ (where the DS \textquoteleft Re\textquoteright$\,$chains are parallel to the transport direction), and \textquoteleft ReS$_{2}$ supercell3\textquoteright$\,$ (where the structure is obtained from the geometry optimized ReS$_{2}$/WSe$_{2}$ and the angle between a$_{1}$-b$_{1}$ and b$_{1}$-c$_{1}$ is $\sim$ 123$^\circ$) \cite{DSahaACSanm,YCLin}.
\\

Figure 7 shows the atomistic models of the two-port devices with \textquoteleft ReS$_{2}$ supercell1\textquoteright$\,$ (length = 4.588 nm and width = 1.135 nm) and \textquoteleft ReS$_{2}$ supercell2\textquoteright$\,$ (length = 4.541 nm and width = 1.311 nm) channel regions. Figure 8 and 9 illustrate the electron density plots and zero bias transmission spectra of \textquoteleft ReS$_{2}$ supercell1\textquoteright$\,$ and \textquoteleft ReS$_{2}$ supercell2\textquoteright$\,$. It is important to realize here that we have purposefully designed those supercell structures where the \textquoteleft Re\textquoteright-\textquoteleft Re\textquoteright$\,$bonds are either parallel or perpendicular to the transport direction (Z direction). However, for single layer distorted 1T ReS$_{2}$, intrinsically the DS \textquoteleft Re\textquoteright$\,$chains form a certain angle with the [100]/[010] axis \cite{DSahaACSanm,YCLin}. Thus, as illustrated in Fig. 10(a), ReS$_{2}$ supercell3\textquoteright$\,$(channel length = 4.620 nm and width = 1.143 nm) essentially exhibits the natural orientation of DS \textquoteleft Re\textquoteright$\,$chains.
\\

Electron density plots of Fig. 8 depict how the valence electrons around \textquoteleft Re\textquoteright$\,$atoms are distributed across the channel regions. Apart from that, significant reduction in energy gap values can be observed for \textquoteleft ReS$_{2}$ supercell1\textquoteright$\,$ and \textquoteleft ReS$_{2}$ supercell2\textquoteright$\,$(Fig. 9), owing to the large transmission states within the gap. Considering the transmission spectra as shown in Fig. 9, distinguishable transmission states within the range of 0.636 eV and 0.920 eV for \textquoteleft ReS$_{2}$ supercell1\textquoteright$\,$ and 0.637 eV and 0.922 eV for \textquoteleft ReS$_{2}$ supercell2\textquoteright, lower the energy gap values ($\sim$ 1.23 eV) of both the supercells.   
\\

However, the calculated energy gap for \textquoteleft ReS$_{2}$ supercell3\textquoteright$\,$(Fig. 10 (b)) is $\sim$ 1.37 eV. More importantly, this value is similar to the bandgap of distorted 1T ReS$_{2}$ unit cell \cite{DSahaACSanm}. It further reinforces that the \textquoteleft ReS$_{2}$ supercell3\textquoteright$\,$has the geometry where orientation of DS \textquoteleft Re\textquoteright$\,$chains as well as distribution of valence electrons around \textquoteleft Re\textquoteright$\,$atoms (Fig. 10(a)), reflects that of the natural single layer. 
\\
\\

\begin{figure}[!htbp]
\begin{center}
\includegraphics [height=1.962in,width=3.2in]{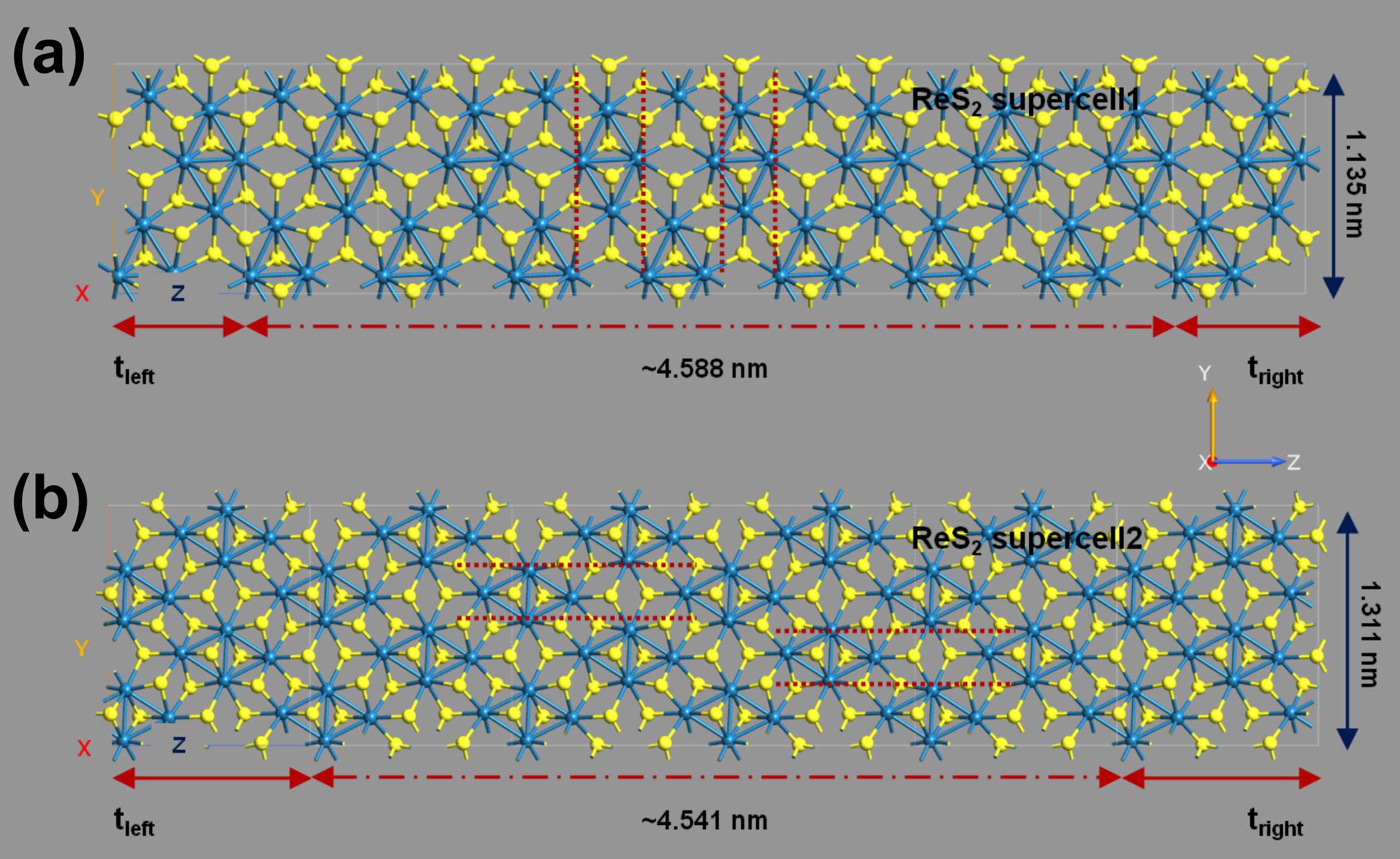}
\caption{Atomistic models of (a) ReS$_{2}$ supercell1 and (b) ReS$_{2}$ supercell2, obtained by keeping the \textquoteleft Re\textquoteright-\textquoteleft Re\textquoteright$\,$bonds are either perpendicular or parallel  to the transport direction.} 
\end{center}
\label{Figure7}
\end{figure}

\begin{figure}[!htbp]
\begin{center}
\includegraphics [height=3.029in,width=3.2in]{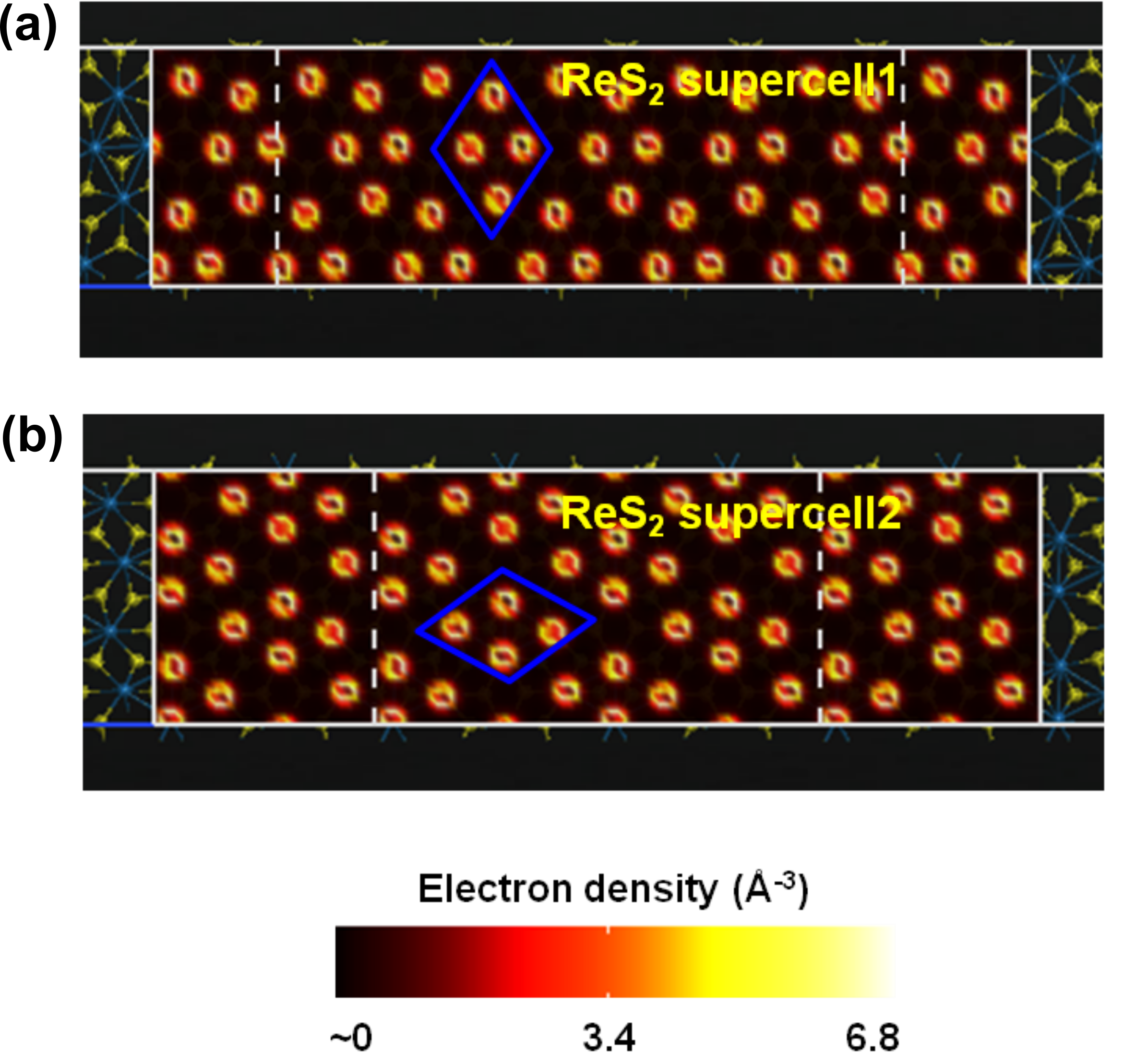}
\caption{Electron density plots depicting the distribution of valence electrons around \textquoteleft Re\textquoteright$\,$atoms of ReS$_{2}$ supercell1 and ReS$_{2}$ supercell2.} 
\end{center}
\label{Figure8}
\end{figure}

\begin{figure}[!htbp]
\begin{center}
\includegraphics [height=1.725in,width=3.4in]{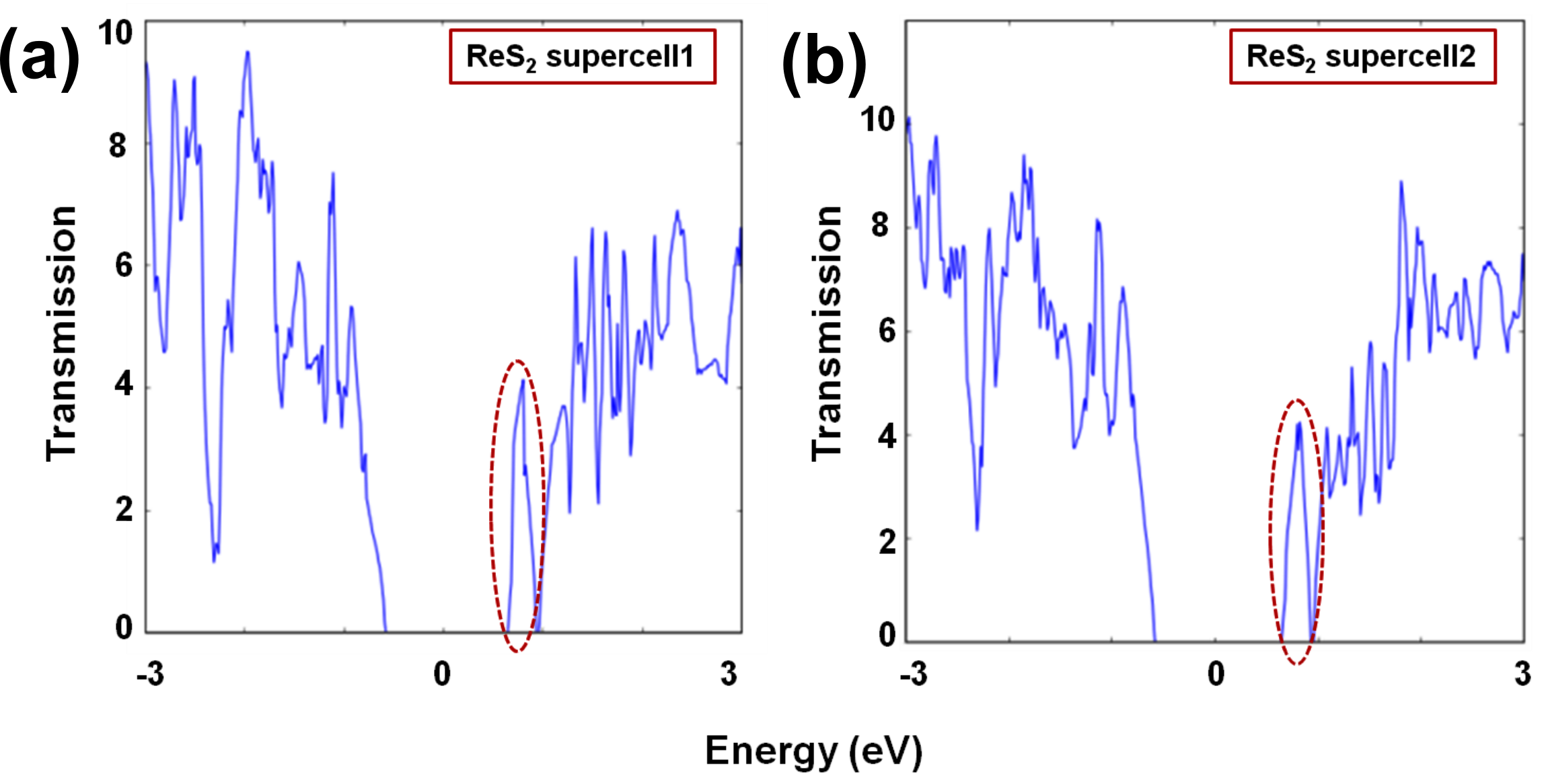}
\caption{Transmission spectra (zero bias) of the two-port devices with (a) ReS$_{2}$ supercell1 and ReS$_{2}$ supercell2 channel regions.} 
\end{center}
\label{Figure9}
\end{figure}

\begin{figure}[!htbp]
\begin{center}
\includegraphics [height=3.43in,width=3in]{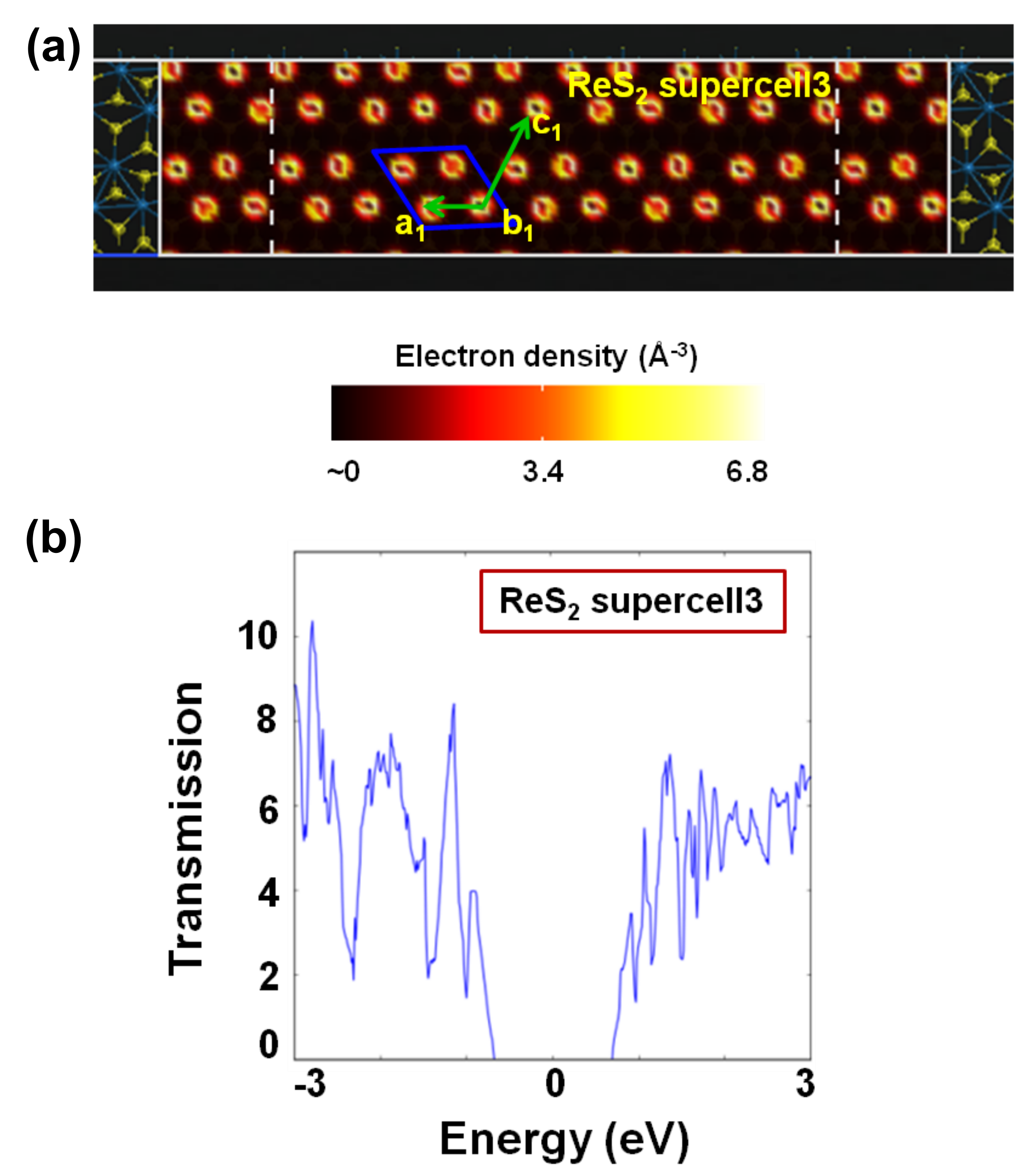}
\caption{Plots illustrating the (a) electron density distribution around \textquoteleft Re\textquoteright$\,$atoms of ReS$_{2}$ supercell3 and (b) zero bias electronic transmission in two-port devices with ReS$_{2}$ supercell3 channel.} 
\end{center}
\label{Figure10}
\end{figure}

\section{Conclusion}

In this work, electronic transmission in single layer distorted 1T ReS$_{2}$ and distorted 1T ReS$_{2}$ based type-II vdW heterointerfaces are studied. We have demonstrated that the ReS$_{2}$/MoSe$_{2}$ heterostructure exhibits stronger effect of interlayer coupling at the heterointerface, owing to the larger average value of charge re-distribution between the constituent layers. Moreover, we have compared the electronic transmission through ReS$_{2}$/WSe$_{2}$ and ReS$_{2}$/MoSe$_{2}$ channels, and computed their near-equilibrium conductance values. We have found that the ReS$_{2}$/MoSe$_{2}$ channel is more conducive to the charge carrier transport. Apart from that, we have explored the in-plane anisotropy of the group-7 constituent material, that is the single layer distorted 1T ReS$_{2}$ and investigated the role of clusterized \textquoteleft Re\textquoteright$\,$ atoms in electronic transmission. This study may further be extended for fewlayer ReS$_{2}$/WSe$_{2}$ and ReS$_{2}$/MoSe$_{2}$ channels, to explore their electronic properties and compare the near-equilibrium conductance values.
\\

\section*{Acknowledgment}
D.S. acknowledges the Department of Electrical Engineering, Indian Institute of Technology (IIT) Bombay for the Institute Post Doctoral Fellowship. Authors also acknowledge support from the Indian Institute of Technology Bombay Nanofabrication Facility (IITBNF). This work was funded by the Department of Science and Technology, Govt. of India through its Swarna Jayanti Fellowship scheme (Grant No. DST/SJF/ETA-01/2016-17).

\onecolumn
\section*{Supporting Information}
\textbf{Band dispersions of the ReS$_{2}$/WSe$_{2}$ and ReS$_{2}$/MoSe$_{2}$ vdW heterointerfaces}
\\
\\
\\

\begin{figure}[!htbp]
\begin{center}
\includegraphics [height=2.715in,width=3.5in]{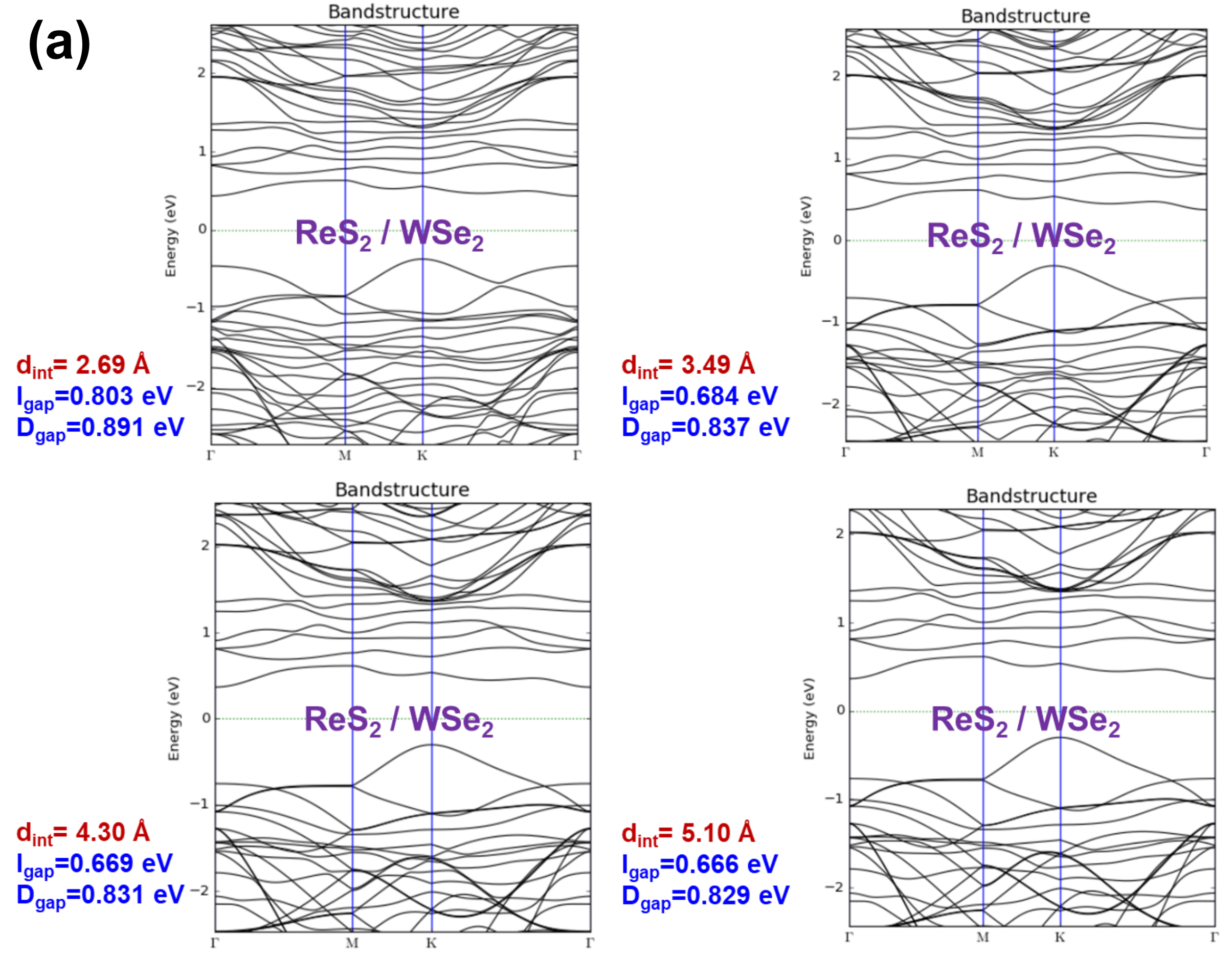}
\caption*{Figure S1 (a) : Band dispersions of the ReS$_{2}$/WSe$_{2}$ heterostructure with the varying interlayer distances.} 
\end{center}
\label{FigureS1a}
\end{figure}

\begin{figure}[!htbp]
\begin{center}
\includegraphics [height=2.659in,width=3.5in]{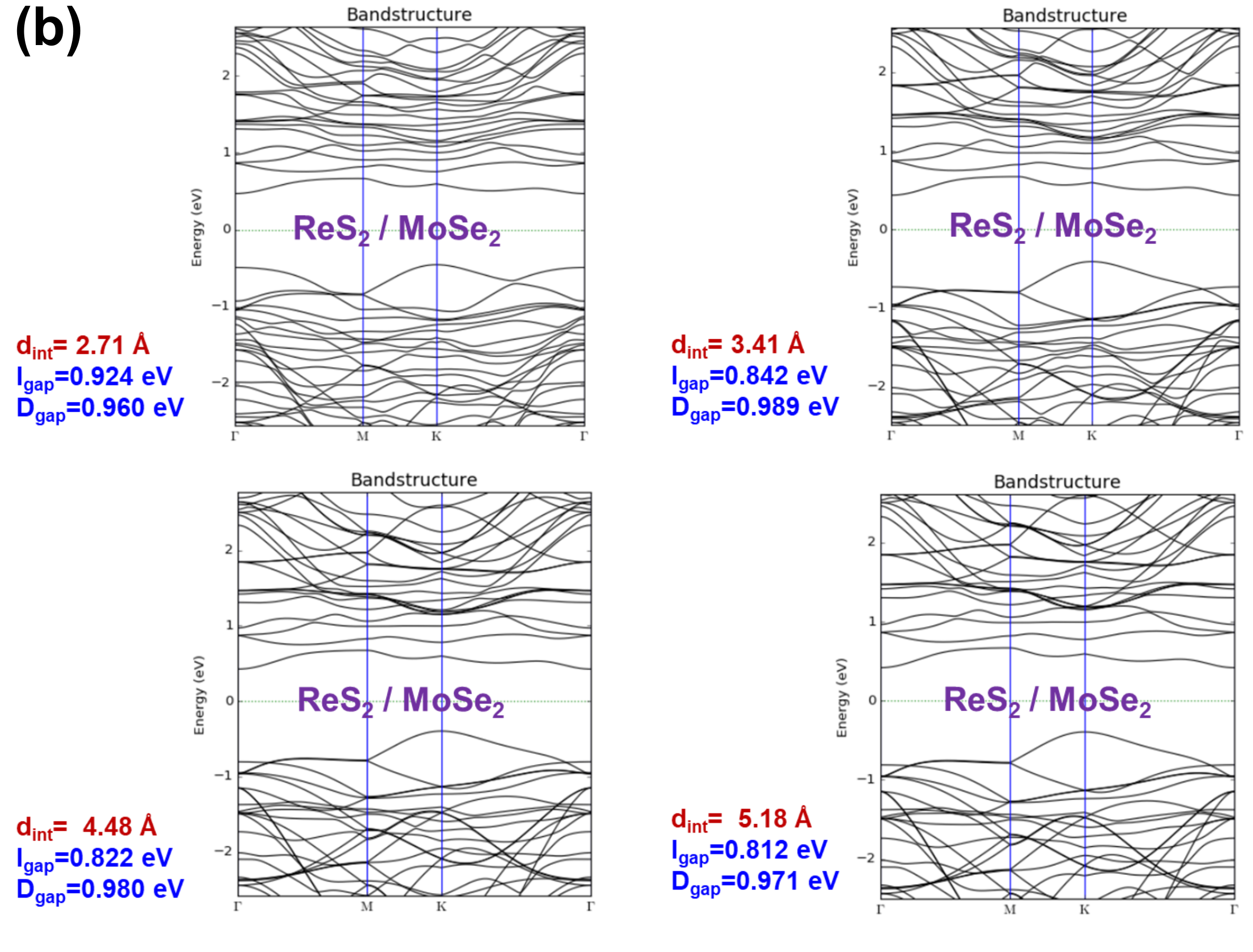}
\caption*{Figure S1 (b) : Band dispersions of the ReS$_{2}$/MoSe$_{2}$ heterostructure with the varying interlayer distances.} 
\end{center}
\label{FigureS1b}
\end{figure}

 \newpage
\textbf{Seebeck coefficients plots (energy range of -3 eV to 3 eV)}
\\
\\
\\

\begin{figure}[!htbp]
\begin{center}
\includegraphics [height=1.58in,width=3.5in]{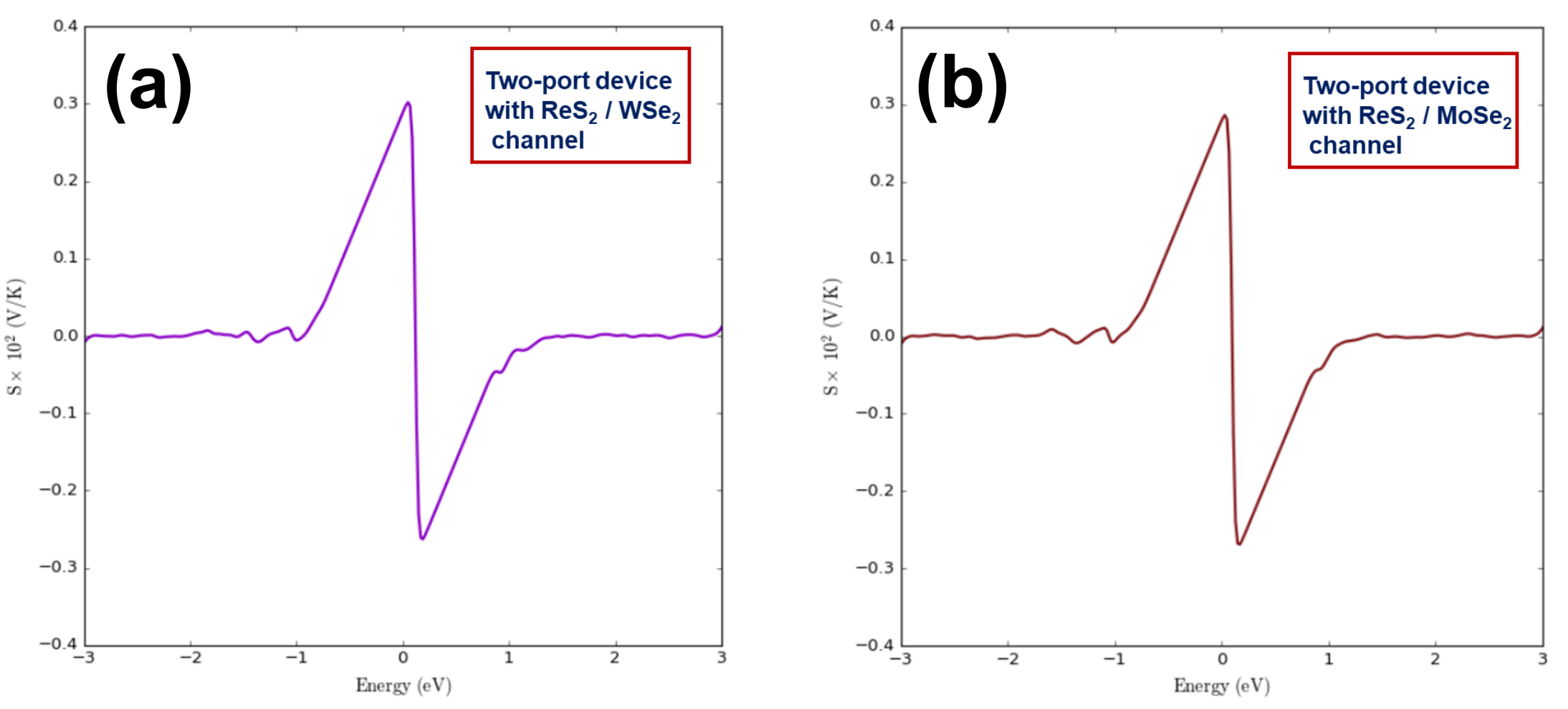}
\caption*{Figure S2 : Seebeck coefficient plots of the two-port devices with (a) ReS$_{2}$/WSe$_{2}$ and (b) ReS$_{2}$/MoSe$_{2}$ channels.} 
\end{center}
\label{FigureS2}
\end{figure}

\end{document}